\documentclass[12pt]{article}

\usepackage{a4wide}
\usepackage{amssymb}
\usepackage{amsthm}
\usepackage{amsmath}

\newcommand{\C}{\mathbb{C}}
\newcommand{\R}{\mathbb{R}}
\newcommand{\N}{\mathbb{N}}
\newcommand{\Z}{\mathbb{Z}}
\newcommand{\E}{\mathbb{E}}
\newcommand{\T}{\mathbb{T}}
\newcommand{\OO}{\mathcal{O}}

\newcommand{\h}{{{\textstyle \frac{1}{2}}}}
\newcommand{\dett}{{\rm det}_2\ }
\newcommand{\Leen}[1]{\mathbb L_1(#1)}

\newcommand{\Ltwee}{{\mathbb{L}_2}}
\newcommand{\Linfty}{{\mathbb{L}_\infty}}

\newcommand{\supp}{\mathop{\mathrm{supp}\,}}
\newcommand{\seq}[3]{\{ #1_{#2}\}_{#2 \in #3}}

\DeclareMathOperator{\Tr}{Tr}
\long\def\symbolfootnote[#1]#2{\begingroup%
\def\thefootnote{\fnsymbol{footnote}}\footnote[#1]{#2}\endgroup}

\newtheorem{theorem}{Theorem}[section]
\newtheorem{lemma}[theorem]{Lemma}
\newtheorem{corollary}[theorem]{Corollary}
\newtheorem{proposition}[theorem]{Proposition}
\newtheorem{Remark}[theorem]{Remark}

\numberwithin{equation}{section} 
\title{Powers of large random unitary matrices and Toeplitz
determinants}
\author{Maurice Duits \thanks{ The  author is a research assistant of the Fund for
Scientific Research -- Flanders and was supported by
  the Marie Curie Training
Network ENIGMA, European Science Foundation Program MISGAM, FWO-Flanders project G.0455.04,
K.U. Leuven research grant OT/04/21 and
Belgian Interuniversity Attraction Pole P06/02
} \and Kurt Johansson \thanks{Supported by the
G\"oran Gustafsson Foundation (KVA). } }

\date{}

\begin{document}
 \maketitle
\abstract{
  We  study the limiting behavior of  $\Tr
  U^{k(n)}$, where $U$ is a $n\times n$ random unitary matrix  and $k(n)$ is a natural number that may vary with $n$ in an arbitrary way. Our analysis is based on the connection
  with Toeplitz determinants.   The central
  observation of this paper is a strong Szeg\"o limit theorem for Toeplitz determinants associated to symbols
  depending on $n$ in a particular way. As a consequence to this
  result, we find that  for each fixed $m\in \N$, the random  variables $ \Tr U^{k_j(n)}/\sqrt{\min(k_j(n),n)}$,  $j=1,\ldots, m$, converge to independent
standard complex normals. }

\section{Introduction and statement of results}

\subsubsection*{Random matrix theory}

Let $U$ be a random unitary matrix with respect to the Haar measure
on $U(n)$, where $U(n)$ is the group of unitary matrices of size
$n\times n$. Denote the eigenvalues of $U$ by $ {\rm e}^{{\rm i}
\theta_\mu }$, for $\mu=1,\ldots,n$ with $\theta_\mu\in [-\pi,\pi)$.
Throughout this paper we will consider the random variable $X_n$ defined by
\begin{equation}
  X_n(U)=\sum_{\mu=1}^n f_n({\rm e}^{{\rm i}\theta_\mu}),
\end{equation}
where $f_n $ is a square integrable function on $\T=\{z\in \C\ : \
|z|=1\}$ with Fourier-series
\begin{equation} \label{symbol}
f_n(z)=\sum_{|j|>0}\frac{\alpha_j z^{k_j(n)}}{\sqrt{\min(|k_j(n)|,n)}} .
\end{equation}
Here we assume that  $\seq{\alpha}{j}{\Z}$ is a square summable
sequence satisfying $\alpha_{-j}=\overline{\alpha_j}$, for each $n
\in \N$ the sequence $\{k_j(n)\}_{j\in \N}$  consists of mutually
distinct positive integers and $k_{-j}(n)=-k_j(n)$. Under these
conditions $f_n$ is real-valued. Alternatively, we may write $X_n$
as
\begin{equation}
X_n(U)=\sum_{|j|>0}\frac{\alpha_j}{\sqrt{\min(|k_j(n)|,n)}} \Tr U^{k_j(n)}.
\end{equation}
The main result we obtain is the following theorem.
\begin{theorem}  \label{limiet charact functie}
We have that
  \begin{equation}
\lim_{n\to \infty} \E[{\rm e}^{{\rm i X_n}}]={\rm
e}^{-\sum_{j=1}^\infty |\alpha_j|^2}.
\end{equation}
Hence, for each fixed $m\in \N$, the random  variables $ \frac{1}{
\sqrt{min(k_j(n),n)}}$ $\Tr U^{k_j(n)}$,  for $j=1,\ldots, m$, converge
to independent standard complex normals.
\end{theorem}

The latter result was obtained before in several special cases. When
$k_j(n)$, $1\leq j \leq m$, do not depend on $n$, this  result is
proved by Diaconis and Shahshahani \cite{DiacShahs} using moment identities.
 In this case it is in fact a  direct consequence of
the strong Szeg\"o limit theorem for Toeplitz determinants and the
Weyl integration formula. If we consider a single $k_j(n)\geq n$
this result is due to Rains, see \cite{Rains}. More details  and an
extensive list of references  can be found in the survey article by
Diaconis \cite{Diac}. See also  Diaconis and Evans \cite{DiacEv}.

There is a remarkable difference in normalization between the two
cases $k_j(n)\leq n$ and $k_j(n)>n$. For the single case $k_j(n)>
n$, Rains proved that the eigenvalues of $U^{k_j(n)}$  behave like
$n$ independently and uniformly distributed points on  the unit
circle. Therefore (\ref{limiet charact functie}) follows from the
classical central limit theorem. In particular, the sum of the
eigenvalues is of order $\sqrt{n}$.

 For $k_j(n) \leq n$, the term $\Tr U^{k_j(n)}$ is normalized by $\sqrt{k_j(n)}$.
This normalization follows from the correlation between the
eigenvalues of $U^{k_j(n)}$. Due to repulsion, the typical picture
one finds for the eigenvalues is that of a small perturbation of $n$
equidistant points on the unit circle and we have a very effective
cancellation. Note that the sum of $n$ equidistant points on the
unit circle is zero.

Our result generalizes previous results by allowing arbitrary powers
depending on $n$ and thus combines the result from Szeg\"o's theorem
with that of Rains.

An interesting generalization of the problem we consider would be to
allow the coefficients $\alpha_j$ to depend on $n$. In this case it
seems difficult to formulate a general theorem. See section
\ref{Generalisatie and Sosh} for a remark.

\subsubsection*{Strong Szeg\"o limit for $n$-dependent symbols}

The starting point of our analysis is the connection with Toeplitz
determinants. If $a\in \Leen{\T}$, let $T_n(a)$ be the $n\times n$
matrix given by $\big(T_n(a)\big)_{jk}=a_{j-k}$, where the $a_k$ are
the Fourier-coefficients of $a$. The Heine-Szeg\"o identity states
that
\begin{equation} \label{connect expect toepl}
 \E[{\rm e}^{{\rm i X_n}}]= \det T_n( {\rm e}^{{\rm i} f_n}),
\end{equation}
see \cite{Diac}. Using this identity we see that in case $k_j(n)$, $1\leq j \leq m$, do not depend
on $n$, Theorem \ref{limiet charact functie} is nothing else then
the strong Szeg\"o limit for Toeplitz determinants. In order to
prove Theorem \ref{limiet charact functie} in the general case, we
will prove a strong Szeg\"o limit for $n$-dependent symbols of the
type
(\ref{symbol}).\\

Note that $f_n$ as defined in (\ref{symbol}) is a real-valued
function. The strong Szeg\"o limit that we prove holds for
complex-valued functions as well, but with  a stronger condition on
the coefficients $\alpha_j$. For the sake of completeness we will
prove the general complex-valued case.

 Let $\seq{\alpha}{j}{\Z}$ be any sequence of complex
 numbers satisfying $\sum_j |\alpha_j|<\infty$. For each $n \in \N$ let  $\{k_j(n)\}_{j\in \N}$ again  be a sequence of mutually distinct positive integers and set  $k_{-j}(n)=-k_j(n)$. Define $g_n:\T\to \C$ by
\begin{equation}
g_n(z)=\sum_{|j|>0}\frac{\alpha_j
z^{k_j(n)}}{\sqrt{\min(|k_j(n)|,n)}} ,
\end{equation}
for all $z\in \T$ and $n\in \N$.  Our main result is the following
\begin{theorem}  \label{main result}  If $\sum_{j} | \alpha_j|<\infty$, then
\begin{equation}
\lim_{n\to \infty}\det T_n({\rm e}^{g_n})=\exp {\sum_{j=1}^\infty \alpha_j\alpha_{-j} }.
\end{equation}
\end{theorem}
This is   the analogue of the strong Szeg\"o  theorem for Toeplitz determinants,
but now for symbols that vary with $n$ in a particular way.\\

Now Theorem \ref{limiet charact functie} follows  from (\ref{connect
expect toepl}) and Theorem \ref{main result} with $g_n={\rm i} f_n$,
but under the extra condition $\sum_{j} |\alpha_j|<\infty$. This
condition can however be eliminated  by a standard approximation
argument which is described in Section \ref{Approximation argument}.
However, we want to emphasize that this argument depends on the fact
that $f_n$ is real-valued.

\subsubsection*{Overview of the proof}

We will omit the dependence on $n$ in the notation and simply write
$g$ and $k_j$. Split $g$ in
\begin{equation} g(z)=g^{(1)}(z)+g^{(2)}(z)=\sum_{0<|k_j|\leq n } \frac{\alpha_j
z^{k_j}}{\sqrt{|k_j|}} +  \sum_{|k_j|> n} \frac{\alpha_j
z^{k_j}}{\sqrt{n}} \end{equation} Let $a$ and $b$ be defined by
\begin{align} \label{definitie a en b}
a={\rm e}^{g^{(1)}} \qquad  \textrm{and} \qquad b= {\rm e}^{g^{(2)}}.
\end{align}
 Define
\begin{align}
C^{(1)}=\sum_{0< k_j \leq n} \alpha_j \alpha_{-j} ,\qquad
C^{(2)}=\sum_{k_j> n} \alpha_j \alpha_{-j},
 \qquad C=\sum_{j =1}^\infty \alpha_j\alpha_{-j},
\end{align}
Note that $C^{(1)}$ and $C^{(2)}$ depend on $n$, whereas $C$  does not.

The terms $a$ and $b$ are very different in behavior. As a
consequence, we  analyze them separately.
 We therefore divide the proof of Theorem \ref{main result}  into two parts. The first part consists of proving that
\begin{equation} \label{main result deel1}
  \lim_{n\to \infty} {\rm e}^{-C^{(1)}} \det T_n(a)=1.
\end{equation}
To this end we need the Fredholm determinant identity for Toeplitz
determinants, which was found by Case and Geronimo \cite{GerCa} and
independently
 by Borodin and Okounkov \cite{BorOk}.

The second part consists of proving that
\begin{equation} \label{mian result deel2}
  \lim_{n\to \infty} \frac{{\rm e}^{-C^{(2)}} \det T_n(a b)}{\det T_n (a)}=1.
\end{equation}
Indeed if we can prove that (\ref{main result deel1}) and (\ref{mian result deel2}) hold, then a simple multiplication of the two gives
\begin{align}
  \lim_{n\to \infty} {\rm e}^{-C^{(1)}-C^{(2)}} \det T_n(ab)=\lim_{n\to \infty} {\rm e}^{- C} \det T_n(ab)=1.
\end{align}
Now, since $C$ does not depend on $n$ we can multiply both sides with
${\rm e}^{ C}$ which proves Theorem \ref{main result}.

For reasons of clarity we will prepare the proof of (\ref{mian
result deel2}) and first prove
\begin{equation} \label{main result deel3}
  \lim_{n\to \infty} {\rm e}^{-C^{(2)}} \det T_n(b)=1.
\end{equation}
The proof of this result follows by a fairly direct computation. The
results of this computation  can be used for proving  (\ref{mian
result deel2}). Hence, in the remaining proof of (\ref{mian result
deel2})  we can restrict ourselves to only those parts that come in
by interaction of $g^{(1)}$ and $g^{(2)}$. In our opinion, it helps
to get a better understanding of the problem. Moreover, combining
(\ref{main result deel1}), (\ref{mian result deel2}) and (\ref{main
result deel3}) we immediately find the following result.
\begin{proposition} \label{separation result} We have that
\begin{equation}
\lim_{n\to \infty} \frac{\det T_n(ab)}{\det T_n(a)\det T_n (b )}=1.
\end{equation}
\end{proposition}
This is a so-called separation theorem. Such results have been often
investigated before, see for example \cite{BasWid2,Wieand}. However,
all the results known thus far use the fact that $H(a)H(\tilde b)$
is of trace class. This is not necessarily true in our case, which
makes Theorem \ref{separation result} an interesting result in its
own right.

\section{Preliminaries}

To fix notation, we recall some definitions of certain operators and Banach algebras we need
later. For a more detailed discussion we refer to \cite{BottSilb}.

For $c\in \Linfty(\T)$, define infinite matrices $T(c)$ and $H(c)$
by
\begin{equation}
T(c)=\left(c_{j-l}\right)_{j,l=1}^\infty \qquad \textrm{and} \qquad
H(c)=\left(c_{j+l-1}\right)_{j,l=1}^\infty ,
\end{equation}
where $c_k$ are the Fourier coefficients of $c$. These matrices
induce bounded operators on $ \ell_2(\N)$. Moreover,
$\|T(c)\|_\infty=\|c\|_{\Linfty}$ and $\|H(c)\|_\infty\leq
\|c\|_\infty$.

Denote with $P_n$ the projection operator on $\ell_2$ that projects
on the subspace of all $x\in \ell_2(\N)$ for which $x_k=0$ for all
$k>n$. Define $Q_n=I-P_n$. Let $W_n:\ell_2(\N)\to \ell_2(\N)$ be the
operator defined by
\begin{equation}
\left(W_n x\right)_k=\left\{\begin{array}{cc} x_{n-k+1}, & 1\leq
k\leq n,\\ 0, &  k>n  \end{array}\right.,
\end{equation}
for all $x\in \ell_2(\N)$. If $c\in \Linfty$, then
\begin{equation} \label{Wn rel ctilde}
W_nT_n(c)W_n=T_n(\tilde c),
\end{equation}
where $\tilde c(z)=c(1/z)$.\\

Next we recall the definition of certain Banach algebras which will
appear frequently in the sequel.

The space $B^{1/2}_2$ consists of all $f\in \Ltwee(\T)$ for which
$\sum_{k} |k| | f_k|^2<\infty,$  equipped with   norm defined by
\begin{equation}
\|f\|_{B^{1/2}_2}^2=\sum_{k}(1+|k|) | f_k|^2.
\end{equation}
Again, the $f_k$ denote the Fourier coefficients of $f$. The space
$B^{1/2}_2$ is a Sobolev space and a Banach algebra.

The Krein algebra $K^{1/2}_2$ is defined as  $B^{1/2}_2 \cap
\Linfty(\T)$. This is a (non-closed) subalgebra of $\Linfty(\T)$.
However, the norm defined by
\begin{equation}
  \|f\|_{K_2^{1/2}} =\|f\|_\Linfty +\|f\|_{B^{1/2}_2},
\end{equation}
for all $f \in K_2^{1/2}$, turns $K^{1/2}_2$ into a Banach algebra.

 The Wiener algebra consists of all $f\in \Linfty$, for which $\sum_k | f_k|<\infty$ and has norm
 \begin{equation}
   \|f\|_W=\sum_{k}| f_k|,
 \end{equation}
 for all $f\in W$. It is  well-known that this is again a Banach algebra.

Note that due to the assumption $\sum |\alpha_j|<\infty$ we have
that $g^{(1)} \in K^{1/2}_2$ and $g^{(2)} \in W$. In particular this
shows that $a$ and $b$ in (\ref{definitie a en b}) are well-defined.
Moreover, $a \in K^{1/2}_2$ , $b\in W$ and we have the following
inequalities
\begin{align}
  \|a\|_{B^{1/2}_2}&\leq {\rm e}^{\|g^{(1)}\|_{B^{1/2}_2}}<{\rm e}^{\left(2 \sum |\alpha_j|^2\right)^{1/2}},
  \label{uniform boundedness of norm a} \\
  \|b\|_{W}&\leq {\rm e}^{\|g^{(2)}\|_{W}}\leq {\rm e}^{\sum
  |\alpha_j|/\sqrt{n}}. \label{uniform boundedness of norm b}
\end{align}
Hence, $\|a\|_{B^{1/2}_2}$ and $\|b\|_W$ are uniformly bounded in
$n$. For convenience we define
\begin{align} \label{definitionA1A2}
  A_1=\sum |\alpha_j| \qquad \textrm{and} \qquad A_2=\left(\sum  |\alpha_j|^2 \right)^{1/2}.
\end{align}
These constants will appear frequently in upcoming inequalities. \\

Besides the operator norm $\|\cdot\|_\infty$ we will also use the
trace norm, denoted by $\|\cdot\|_1$,  and the Hilbert-Schmidt norm,
denoted by $\|\cdot \|_2$. Note that if $c\in K^{1/2}_2$, then
$H(c)$ is a Hilbert-Schmidt operator and
\begin{equation}
\|H(c)\|_2^2=\sum_{j,l=1}^\infty |c_{j+l-1} |^2=\sum_{j=1}^\infty
j|c_j|^2 \leq \|c\|_{B_2^{1/2}}^2.
\end{equation}
This will be used frequently in the sequel.

\section{Proof of Theorem \ref{main result}}

\subsection{Proof of \eqref{main result deel1}}

First, we will prove (\ref{main result deel1}). To this end we will
use a celebrated Fredholm identity for Toeplitz determinants. Let
$g_+^{(1)}$ be the projection of $g^{(1)}$ onto the subspace of all
$ f\in K^{1/2}_2 $ for which $ f_k=0$ for all $k<0$. Moreover,
define $g^{(1)}_-=g^{(1)}-g^{ (1)}_+$, $a_+={\rm e}^{g^{(1)}_+}$ and
$a_- ={\rm e}^{g^{(1)}_-}$. Finally, define $\phi=a_+^{-1}a_-$ and
$\psi=\widetilde{ a_+} \widetilde {a_-^{-1}}$.

The Borodin-Okounkov-Geronimo-Case identity now states that
\begin{equation} \label{BoOk}
\det T_n(a)={\rm e}^{C^{(1)}} \det (I-Q_n H(\phi)H(\psi)Q_n),
\end{equation}
for all $n\in \N$.  Note that since $K_2^{1/2}$ is a Banach algebra,
we find that $\phi,\psi \in K_2^{1/2}$ and hence $ Q_n
H(\phi)H(\psi)Q_n$ is a trace class operator. The  determinant on
the right-hand side is a Fredholm-determinant.  Note that we use the
formulation by Basor and Widom, see \cite{BasWid}, which is slightly
different from the one by Borodin and Okounkov in \cite{BorOk}.

So we need to prove that the Fredholm-determinant converges to 1 to
obtain  (\ref{main result deel1}).
\begin{lemma} \label{lemma: estimates on Fredhdet} We have that
\begin{equation}
|\det (I-Q_n H(\phi)H(\psi)Q_n)-1|\leq \exp \left(\left(\sum_{k=1}^\infty k | \phi_{k+n}|^2\right)^{1/2} \left( \sum_{k=1}^\infty k |\psi_{k+n}|^2\right)^{1/2}\right)-1,
\end{equation}
for all $n\in \mathbb{N}$.
\end{lemma}
\begin{proof}
A standard inequality for  Fredholm-determinants gives
\[|\det (I-Q_n H(\phi)H(\psi)Q_n)-1|\leq {\rm e}^{\|Q_n H(\phi)H(\psi)Q_n\|_1 }-1.\]
The trace norm can be estimated by
 \[\|Q_n H(\phi)H(\psi)Q_n\|_1 \leq \| Q_n H(\phi)\|_2\|H(\psi)Q_n\|_2.\]   A straightforward calculation shows that
\[\|Q_n H(\phi)\|^2_2=\sum_{k=1}^\infty k | \phi_{k+n}|^2, \qquad \textrm{and} \qquad \| H(\psi)Q_n\|^2_2=\sum_{k=1}^\infty k | \psi_{k+n}|^2, \]
which proves the statement.
\end{proof}
Hence we need to show that
\begin{equation}
 \lim_{n \to \infty}\sum_{k=1}^\infty k | \phi_{k+n}|^2= 0 \qquad \textrm{and} \qquad \lim_{n \to \infty}\sum_{k=1}^\infty k | \psi_{k+n}|^2=0.
\end{equation}
Note that if $\phi$ and $\psi$ did not depend on $n$ (as in the classical case), then this trivially holds. But since they depend on $n$ there is still some work to be done.

\begin{lemma} \label{estimate fourier coeff 1} Let $N \in \mathbb{N}$ and $t$ be defined by the Fourier series
$  t(z)=\sum_{0<j\leq N} \frac{t_j z^j}{\sqrt{|j|} }$. Define $F_t$
associated to $t$ by $F_t(z)=\sum_{0<j\leq N} |t_j| z^{j}.$ Then
\begin{equation}
|({\rm e} ^t)_{k+N}|<\frac{1}{\sqrt{k(N+k)}} \big(F_t({\rm e}^{F_t}-1)\big)_{N+k}
\end{equation}
for all $k\in \N$.
\end{lemma}
\begin{proof}
First consider powers  $t^l$ for $l\geq 2$. Then
\begin{equation*}
\big(t^l\big)_{k+N}=\sum_{j_1+j_2+\cdots + j_l=k+N}
\frac{t_{j_1}\cdots t_{j_l}}{\sqrt{|j_1\cdots {j_l}|}}
\end{equation*}
Since ${j_1}+{j_2}+\cdots + {j_l}=k+N$, there should be at least one
${j_s}$, with $j_{s} \geq (N+k)/l$. But ${j_s} \leq N$ and hence
${j_1}+{j_2}+\cdots + {j_l}-{j_s}\geq k$. Hence there exists a
${j_r} \neq {j_s}$ such that ${j_r} \geq k/(l-1)>k/l$.

Therefore
\begin{align*}
\Big|\big({t^l}\big)_{k+N}\Big|&< \frac{l}{\sqrt{k(k+N)}}
\sum_{{j_1}+{j_2}+\cdots + {j_l}=k+N} {|t_{j_1}\cdots t_{j_l}|}=
\frac{l}{\sqrt{k(k+N)}} \big({ F_t^l}\big)_{k+N}.
\end{align*}
Hence,
\begin{align*}
\Big|\big({{\rm e}^{t}}\big)_{k+N}\Big|&\leq \sum_{l=2}^{\infty}\frac{\Big|\big( {t^l}\big)_{k+N}\big|}{l!}<\sum_{l=2}^{\infty}\frac{ \big({ F_t^l}\big)_{k+N}}{(l-1)!\sqrt{k(k+N)}}  =\frac{1}{\sqrt{k(k+N)}} \big(F_t({\rm e}^{F_t}-1)\big)_{k+N}
\end{align*}
This proves the statement.
\end{proof}
Now we immediately find the following corollary.
\begin{corollary} \label{estimates Fourier coeff 2}  With $A_1$ as in \eqref{definitionA1A2} we have that
\begin{equation}
\sum_{k=1}^\infty k |\phi_{k+n}|^2< \frac{A_1({\rm e}^{A_1}-1)}{n}
\end{equation}
for all $n$. The same estimate holds for $\psi$.
\end{corollary}
\begin{proof}
Applying Lemma \ref{estimate fourier coeff 1} with $t=\phi$ and $N=n$, we find
\[\sum_{k=1}^\infty k |\phi_{k+n}|^2< \frac{\|F_\phi ({\rm e}^{F_\phi}-1)\|_{\Ltwee}^2}{n}.\]
The statement now follows from the fact that $\|\cdot \|_{\Ltwee}\leq \|\cdot \|_W$, the fact that $W$ is a Banach algebra and $\|F_\phi\|_W\leq A_1$.
\end{proof}
Now (\ref{main result deel1}) follows by combining Corollary
\ref{estimates Fourier coeff 2}, Lemma \ref{lemma: estimates on
Fredhdet} and \eqref{BoOk}.

\subsection{Proof of \eqref{main result deel3}}

Next we analyze $\det T_n(b)$.  In this case the identity
(\ref{BoOk}) breaks down at two places. First,  the factor in front
of the Fredholm-determinant is infinite, since $b$ is not
necessarily contained in $K^{1/2}_2$. Second, the operator in the
Fredholm-determinant is no longer  of trace class and the
determinant is therefore not well-defined. However, there is no need
for such a strong result as (\ref{BoOk}), since a direct  analysis
on $\det T_n(b)$ will suffice.

We will use the notion of regularized determinants. For a trace class operator $A$ the regularized determinant  is defined by
\begin{equation}
  \dett(I+A)={\rm e}^{-\Tr A}\det(I+A).
\end{equation}
One can prove that $A \mapsto \dett(I+A)$  is a continuous function
defined on a dense subspace (namely the space of all trace class
operators) of the space of Hilbert-Schmidt operators. Therefore it
can  be extended and defined for all Hilbert-Schmidt operators.
Moreover, we have that
\begin{equation}  \label{ineq reg det}
  |\dett(I+A)-1|\leq \|A\|_2 \exp\left(\h(\|A\|_2+1)^2\right),
\end{equation}
for all Hilbert-Schmidt operators.

We will use the regularized determinant only for matrices, but  (\ref{ineq reg det}) plays a crucial role.
Write
\begin{align} \label{detnaardett}
  \det T_n(b)&=\det(I+T_n(b-1))={\rm e}^{\Tr T_n(b-1)}\dett(I+T_n(b-1)).
\end{align}
The proof of (\ref{main result deel3}) falls into two parts. First
we will show that the Hilbert-Schmidt norm of $T_n(b-1)$ tends to
$0$ as  $n\to \infty $, hence the regularized determinant tends to
1. And second, we show that $\Tr T_n(b-1)-C^{(2)}$ tends to $0$ as $n\to
\infty$. Then (\ref{main result deel3}) follows by (\ref{ineq reg
det}) and (\ref{detnaardett}).

We start with the trace of $T_n(b-1)$. We define $g^{(2)}_+$ as
$\sum_{|k_j|>n} \frac{\alpha_j z^{k_j}}{\sqrt{n}}$ and
$g^{(2)}_-=g^{(2)}-g^{(2)}_+$. Moreover, we let $b_\pm={\rm
e}^{{g^{(2)}_\pm}}$.

\begin{lemma} \label{trace geval g1is0} With $A_1$ as in \eqref{definitionA1A2} we have that
  \begin{equation}
    \left|\Tr T_n(b-1)-C^{(2)} \right|\leq  { n ({\rm e}^{A_1/\sqrt{n}}-1)^2-A_1^2},
  \end{equation}
  for all $n \in \N$.
\end{lemma}
\begin{proof}
First note that $\Tr T_n(b-1)=n(b_0-1)$. Now

\begin{align*}
  b_0-1&=\sum_{j\geq 0}\left({\rm e}^{ g^{(2)}_+}\right)_j \left({\rm e}^{ g^{(2)}_-}\right)_{-j} -1=\sum_{j\geq 0}\left({\rm e}^{ g^{(2)}_+}-1\right)_j \left({\rm e}^{ g^{(2)}_-}-1\right)_{-j} \\
&=\sum_{l=1}^\infty \sum_{m=1}^\infty \sum_{j\geq 0}\frac{\left({ g^{(2)}_+}^l\right)_j  \left({ g^{(2)}_-}^m\right)_{-j}}{l!m!}
\end{align*} Since $\sum_{j >n} \alpha_j \alpha_{-j}=n\sum_{j\geq 0 }\left( g^{(2)}_+\right)_j \left( g^{(2)}_-\right)_{-j}$, we find

\begin{align*}
\frac{1}{n}\Big|\Tr T_n(b-1) &-\sum_{j >n} \alpha_j \alpha_{-j}\Big| = \left|b_0-1-\sum_{j\geq 0 }\left( g^{(2)}_+\right)_j  \left( g^{(2)}_-\right)_{-j}\right|\\
&=\left|\sum_{l=1}^\infty \sum_{m=1}^\infty \sum_{j\geq 0}\frac{\left({ g^{(2)}_+}^l\right)_j  \left({ g^{(2)}_-}^m\right)_{-j}}{l!m!} -\sum_{j\geq 0 }\left( g^{(2)}_+\right)_j \left( g^{(2)}_-\right)_{-j}\right|
\end{align*}
Now apply the Cauchy-Schwarz inequality  to obtain
\begin{align*}
\frac{1}{n}\Big|\Tr T_n(b-1) &-\sum_{j >n} \alpha_j \alpha_{-j}\Big| \leq \sum_{l=1}^\infty \sum_{m=1}^\infty \frac{\|{ g^{(2)}_+}^l\|_{\Ltwee}  \|{ g^{(2)}_-}^m\|_{\Ltwee}}{l!m!} -\| g^{(2)}_+\|_{\Ltwee} \| g^{(2)}_-\|_{\Ltwee}\\ & \leq \sum_{l=1}^\infty \sum_{m=1}^\infty \frac{\|{ g^{(2)}_+}^l\|_{W}  \|{ g^{(2)}_-}^m\|_{W}}{l!m!} -\| g^{(2)}_+\|_{W} \| g^{(2)}_-\|_{W}\\
&\leq \sum_{l=1}^\infty \sum_{m=1}^\infty \frac{\|{ g^{(2)}_+}\|_{W}^l  \|{ g^{(2)}_-}\|_{W}^m }{l!m!} -\| g^{(2)}_+\|_{W} \| g^{(2)}_-\|_{W}
\end{align*}
Now $\|g_{\pm}^{(2)}\|_W \leq  A_1/\sqrt{n}$ proves the statement.
\end{proof}
Next we proceed with the Hilbert-Schmidt norm of $T_n(b-1)$.
\begin{lemma} \label{hsnorm geval g1is0}  With $A_1$ as in \eqref{definitionA1A2}, we have that
\begin{equation}
  \|T_n(b-1)\|_2\leq \sqrt{n}({\rm e}^{A_1/\sqrt{n}}-1)^2,
\end{equation}
for all $n \in \mathbb{N}$.
\end{lemma}
\begin{proof}
Since $({b_\pm-1})_j=0$ for $j=-n+1,\ldots, n-1$ we find
\begin{align*}
 \|T_n(b-1)\|_2^2 & \leq n \sum_{j=-n+1}^{n-1}| ({b-1})_j|^2 = n \sum_{j=-n+1}^{n-1}| ({ (b_+-1)(b_--1)})_j|^2\\
&\leq n \|(b_+-1)(b_--1)\|_{\Ltwee}^2\leq  n \|(b_+-1)(b_--1)\|_{W}^2\\
&\leq n \|b_+-1\|_W^2 \|b_--1\|_W^2 \leq n ({\rm e}^{\|g^{(2)}_+\|_W}-1)^2  ({\rm e}^{\|g^{(2)}_-\|_W}-1)^2 .
\end{align*}
By $\|g_\pm^{2}\|_W\leq A_1/\sqrt{n}$  we obtain the statement.
\end{proof}
By  Lemma \ref{hsnorm geval g1is0} and \eqref{ineq reg det} we
obtain \begin{equation}   \lim_{n\to \infty} \dett(I+ T_n(b-1))=1.
\end{equation}
By substituting this in \eqref{detnaardett} and using  Lemma
\ref{trace geval g1is0} we obtain \eqref{main result deel3}.

\subsection{Proof of \eqref{mian result deel2}}

Since we proved the result for the cases (\ref{main result deel1})
and (\ref{main result deel3}) in a completely different way,   a
natural way to deal with the general case is to split the two cases.
To this end we use a factorization theorem due to Widom
\begin{equation} \label{Widom fact}
  T_n(ab)=T_n(a)T_n(b)+P_n H(a) H(\widetilde{b}) P_n + W_n H(\widetilde{a}) H(b) W_n,
\end{equation}
and the  operator $B_n$ defined by
\begin{equation} \label{approx inverse}
  B_n=T_n(a^{-1})-P_n H(a_+^{-1})H(\widetilde{a_-^{-1}}) P_n -W_n H(\widetilde{a_-^{-1}}) H(a_+^{-1})W_n.
\end{equation}
The operator $B_n$ is a good approximation of the inverse of
$T_n(a)$. In the case that $a$ does not depend on $n$, this
observation is due to Widom. Moreover, the operator
can be used to prove the strong Szeg\"o limit, see
\cite{BottSilb,Wid}. We will prove that it is also a good
approximation  in our case. One can show, see \cite{BottSilb,Wid},
that
\begin{equation}
B_n T_n(a)=I+P_n H(a_+^{-1})H(\widetilde{a_-^{-1}})Q_n T(a) P_n +W_nH(\widetilde{a_-^{-1}})H(a_+^{-1})Q_n T(\tilde{a})W_n
\end{equation}
for all $n\in \N$. Even in our case where $a$ depends on $n$, the
operators on the right-hand side are small in trace norm.
\begin{lemma} \label{Asymptotic inverse} We have
\begin{equation}\|B_n T_n(a)-I\|_1=\OO(n^{-\h}),
\end{equation}
for $n\to \infty$.
\end{lemma}
\begin{proof}
  First note that
  \begin{align*}
    \|P_n H(a_+^{-1})H(\widetilde{a_-^{-1}})Q_n T(a) P_n\|_1\leq \|P_n H(a_+^{-1})\|_2\|H(\widetilde{a_-^{-1}})Q_n \|_2\| T(a) P_n\|_\infty.
  \end{align*}
Now
\begin{align*}
  \|P_n H(a_+^{-1})\|_2&\leq \|H(a_+^{-1})\|_2\leq \|a_+^{-1}\|_{B^{1/2}_2}< \exp(\sqrt{2} A_2),
  \end{align*}
  and \begin{align*}
  \|T(a)P_n\|_\infty& \leq \|a\|_\infty\leq \|a\|_W \leq \exp(A_1),
  \end{align*}
  and finally
  \begin{align*}
  \|H(\widetilde {a_-^{-1}})Q_n\|_2^2&=\sum_{k=1}^\infty k |(\widetilde{ a_-^{-1}})_{k+n}|^2.
\end{align*}
By Lemma \ref{estimate fourier coeff 1} and the same arguments as in Corollary \ref{estimates Fourier coeff 2},
the latter is  $\OO(n^{-1})$, as $n\to \infty$. This proves the statement.
\end{proof}
Therefore the following corollary is immediate.
\begin{corollary} \label{splitting part 1}  We have that
  \begin{equation}
    \lim_{n\to \infty} \det B_n T_n(a)  =1.
  \end{equation}
\end{corollary}
In view of this corollary,  it is enough to show that
\begin{equation} \label{splitting part 2}
    \lim_{n\to \infty} \exp (- C^{(2)})\det {B_n T_n(ab)}=1,
  \end{equation}
to prove (\ref{mian result deel2}). This will cover the rest of this section.

We will again use the regularized determinant. Write
\begin{equation} \label{BnTnab determinant}
\det B_n T_n(ab)={\rm e}^{\Tr (B_nT_n(ab)-I)} \dett B_n T_n(ab).
\end{equation}
In view of \eqref{detnaardett} and \eqref{ineq reg det},  to prove
(\ref{splitting part 2}) it is enough to (1) prove that $B_n
T_n(ab)-I$ converges to zero in Hilbert-Schmidt norm and (2)
calculate its trace.

If we introduce the notations
\begin{equation}
E_n=-P_n H(a_+^{-1})H(\widetilde{a_-^{-1}}) P_n -W_n H(\widetilde{a_-^{-1}}) H(a_+^{-1})W_n,
\end{equation}
and
\begin{equation}
F_n=P_n H(a) H(\widetilde{b}) P_n + W_n H(\widetilde{a}) H(b) W_n,
\end{equation}
and multiply (\ref{Widom fact}) from the left with $B_n$ we find by (\ref{approx inverse})
\begin{equation} \label{BnTnab ontbonden}
B_nT_n(ab)=B_n T_n(a)T_n(b)+T_n(a^{-1}) F_n +E_n F_n.
\end{equation}
We will analyze the three terms on the right-hand side separately.
In the following lemma, we state results about the Hilbert-Schmidt
norms and the trace of each of these three terms, except for the
trace of $ T_n(a^{-1})F_n$. All the statements follow from earlier
results. However,  $\Tr T_n(a^{-1})F_n$ is more subtle and needs
some extra attention.

\begin{lemma} We have that
\begin{enumerate}
\item $ \|B_n T_n(a)T_n(b)-I\|_2\to 0,$
\item $|\Tr( B_n T_n(a)T_n(b)-I)-C^{(2)}| \to 0,$
\item $\|E_nF_n \|_1\to 0,$
\item $\|T_n(a^{-1})F_n\|_2\to 0,$
\end{enumerate}
for $n\to \infty$.
\end{lemma}
\begin{proof}
  \begin{enumerate}
    \item We estimate the Hilbert-Schmidt norm by \begin{align*}\|B_n T_n(a)T_n(b)-I\|_2& \leq \|\Big(B_n
    T_n(a)-I\Big)T_n(b)\|_2+\|T_n(b-1)\|_2\\
   & \leq \|B_n
    T_n(a)-I\|_2\|T_n(b)\|_\infty+\|T_n(b-1)\|_2.
    \end{align*}
    Note that $\|T_n(b)\|_\infty\leq \|b\|_{\Linfty} \leq \|b\|_W$. The statement now follows  from Lemma \ref{hsnorm geval g1is0}, Lemma \ref{Asymptotic
    inverse} and \eqref{uniform boundedness of norm b}.
\item Note that
\begin{align*}
|\Tr (B_n & T_n(a)T_n(b)-I)-C^{(2)}|\\
&\leq  |\Tr \left(\left(B_n
T_n(a)-I\right)T_n(b)\right)|+|\Tr T_n(b-1)-C^{(2)}|\\
&  \leq  \|\left(B_n T_n(a)-I\right)\|_1\|T_n(b)\|_\infty+|\Tr
T_n(b-1)-C^{(2)}|.
\end{align*} The statement now follows from Lemma
\ref{trace geval g1is0} and Lemma \ref{Asymptotic inverse}.
\item First note that $\|E_n F_n\|_1\leq \|E_n \|_2 \|F_n\|_2$. Now
\begin{align}
  \|F_n\|_2&\leq \|P_n H(a)\|_2 \|P_n H(\tilde b)\|_\infty +\|W_n H(\tilde a)\|_2\|H(b) W_n \|_\infty \nonumber \\
  & \leq \|a\|_{B^{1/2}_2} \|b-1\|_\infty\leq  \|a\|_{B^{1/2}_2} \|b-1\|_W \nonumber \\&\leq \|a\|_{B^{1/2}_2} \Big(\exp(
  A_1/\sqrt{n})-1\Big), \label{HSnorm Fn}
\end{align}
with $A_1$ as in \eqref{definitionA1A2}. By combining \eqref{HSnorm
Fn} with \eqref{uniform boundedness of norm a} we obtain
$\|F_n\|_2\to 0$. By similar estimates one finds that $\|E_n\|_2$ is
bounded in $n$.
\item This follows from \eqref{HSnorm Fn} and the estimate $\|T_n(a^{-1})F_n\|_2\leq \|T_n(a^{-1})\|_\infty \|F_n\|_2$. Note that $\|T_n(a^{-1})\|_\infty \leq \|a^{-1}\|_\infty \leq \|a^{-1}\|_W$ and the latter is uniformly bounded in $n$.
  \end{enumerate}
\end{proof}
From this lemma,  (\ref{BnTnab determinant}), (\ref{ineq reg det})
and (\ref{BnTnab ontbonden}) it follows that
\begin{equation} \label{eq:towardsfinal}
 \lim_{n\to \infty} \exp\Big(- C-\Tr(T_n(a^{-1}) F_n)\Big)\det B_n T_n(ab)=1.
\end{equation}
 Hence it remains to prove that  $\Tr T_n(a^{-1})F_n$ tends to $0$
as $n \to \infty$, which is the most difficult part of the proof. We
start with an estimate that follows from a subtle cancellation.

\begin{lemma} \label{cancellation lemma} There exists a constant $D$ such that
\begin{equation} \label{eq: cancellation lemma}
\sum_{s=-\sqrt{n}}^{\sqrt{n}} \big({a^{-1}}\big)_{s}
\big({a}\big)_{N-s}\leq \frac{D}{ n^{3/4}},
\end{equation}
 for all $n,N\in \N$ with $N>n$.
\end{lemma}
\begin{proof} Let $n,N\in \N$ with $N>n$. Define $j^*=\sup\{j\ | \  k_j<n\}.$
The proof follows by an induction-like argument with respect to
$j^{*}$.

Suppose first that  $k_{j^*}$ is  such that  $N-k_{j^*}>\sqrt{n}/2$. In this case split the sum into two parts
\begin{align} \label{cancel:lemma:splitting1}
\sum_{s=-\sqrt{n}}^{\sqrt{n}} \big({a^{-1}}\big)_{s}
\big({a}\big)_{N-s}=\sum_{|s|<\sqrt{n}/3} \big({a^{-1}}\big)_{s}
\big({a}\big)_{N-s}+\sum_{\sqrt{n}/3 \leq |s| \leq
\sqrt{n}}\big({a^{-1}}\big)_{s}  \big({a}\big)_{N-s}
\end{align}
The second sum of the right-hand side of
\eqref{cancel:lemma:splitting1} is estimated by
\begin{align}
  \left|\sum_{\sqrt{n}/3<|s|<\sqrt{n}} \big({a^{-1}}\big)_{s}  \big({a}\big)_{N-s} \right|
  &\leq \left(\sum_{\sqrt{n}/3<|s|<\sqrt{n}} |\big({a^{-1}}\big)_{s}|^2 \right)^{1/2}\left(\sum_{\sqrt{n}/3 <|s|<\sqrt{n}} |\big({a}\big)_{N-s}|^2\right)^{1/2}\nonumber\\
&\hspace{-2cm}=\left(\sum_{\sqrt{n}/3<|s|<\sqrt{n}} \frac{|s|\, |\big({a^{-1}}\big)_{s}|^2 }{|s|} \right)^{1/2}\left(\sum_{\sqrt{n}/3 <|s|<\sqrt{n}} \frac{|N-s| \, |\big({a}\big)_{N-s}|^2}{|N-s|}\right)^{1/2}\nonumber\\
&\hspace*{-2cm}\leq \frac{\sqrt{3} \|a^{-1}\|_{B^{1/2}_2}
\|a\|_{B^{1/2}_2}}{n^{1/4}\sqrt{N-\sqrt{n}}}\leq \frac{\sqrt{6}
\|a^{-1}\|_{B^{1/2}_2} \|a\|_{B^{1/2}_2}}{n^{3/4}},
\label{cancel:lemma:splitting1:part1}
\end{align}
where we used that $N-\sqrt{n}\geq n+1-\sqrt{n}>n/2$. Note that
$\|a^{-1}\|_{B^{1/2}_2}$ and $\|a\|_{B^{1/2}_2}$ are uniformly
bounded in $n$ by \eqref{uniform boundedness of norm a}.

 The first sum of the right-hand side of
 \eqref{cancel:lemma:splitting1} is estimated in a similar way
\begin{align*}
  \left|\sum_{|s|<\sqrt{n}/3} \big({a^{-1}}\big)_{s}  ({a})_{N-s} \right|\leq \|a^{-1}\|_{\Ltwee}\left(\sum_{|s|<\sqrt{n}/3} |({a})_{N-s}|^2\right)^{1/2}.
\end{align*}
The term $\|a^{-1}\|_\Ltwee$ is uniformly bounded in $n$. Applying
Lemma \ref{estimate fourier coeff 1}, with $t=a$, gives
\begin{align}
  \sum_{|s|<\sqrt{n}/3} |({a})_{N-s}|^2&<\sum_{|s|<\sqrt{n}/3}
  \frac{ |\big(F_a({\rm e}^{F_a}-1)\big)_{N-s}|^2}{(N-s-k_{j^*})(N-s)}\nonumber\\
  &\leq \frac{1}{(N-\sqrt{n}/3-k_{j^*})(N-\sqrt{n}/3)} \sum_{s<\sqrt{n}/3}
  |\big(F_a({\rm e}^{F_a}-1)\big)_{N-s}|^2\nonumber\\
 & \leq \frac{18}{n^{3/2}} \|\big(F_a({\rm e}^{F_a}-1)\|_\Ltwee\leq \frac{18}{n^{3/2}} \|\big(F_a({\rm e}^{F_a}-1)\|_W
 \nonumber \\
 &\leq \frac{18}{n^{3/2}} A_1({\rm
 e}^{A_1}-1).  \label{cancel:lemma:splitting1:part2}
\end{align}
By combining \eqref{cancel:lemma:splitting1},
\eqref{cancel:lemma:splitting1:part1} and
\eqref{cancel:lemma:splitting1:part2}
we obtain the statement in the case $N-k_{j^*}>\sqrt{n}/2$. \\

Now suppose $N-k_{j^{*}}\leq\sqrt{n}/2$. We will then show that the terms that come from $j^{*}$ are negligible. To be precise, define
\begin{align}c_1&= \exp\big((\alpha_{j^*} z^{k_j^{*}} +\alpha_{-j^*}z^{-k_{j^*}})/\sqrt{k_{j^*}}\big), \label{inductie stap in cancelletion lemma 1} \\
a_1&=a c_1^{-1}\label{inductie stap in cancelletion lemma 2}.\end{align}
 We will show that
\begin{align} \label{inductie stap main}
 \left| \sum_{s=-\sqrt{n}}^{\sqrt{n}} \big({a^{-1}}\big)_{s}  \big({a}\big)_{N-s}-\sum_{s=-\sqrt{n}}^{\sqrt{n}} \big({a^{-1}_1}\big)_{s}
 \big({a_1}\big)_{N-s}\right|\leq
 (|\alpha_{j^{*}}|+ |\alpha_{-j^*}|) D_1/n,
\end{align}
where $D_1$ is a constant independent of $j^{*}$, $n$ and $N$ that
can be expressed in terms of $A_1$ and $A_2$ only. Redefine $j^*$,
now with respect to $a_1$. If $N-k_{j^*}>\sqrt{n}/2$, then the above
arguments show that  \eqref{eq: cancellation lemma} holds for $a_1$.
By combining this with \eqref{inductie stap main} we see that
\eqref{eq: cancellation lemma} also holds for $a$.  If however
$N-k_{j^*}<\sqrt{n}/2$ then we  define $a_2$ and $c_2$ as in
(\ref{inductie stap in cancelletion lemma 1}) and (\ref{inductie
stap in cancelletion lemma 2}) and redefine $j^*$ with respect to
$a_1$. We also have that the inequality \eqref{inductie stap main} holds with $a_1$ replaced by $a_2$, $a$
replaced by $a_1$ and $j^*$ is with respect to $a_1$.
If $N-k_{j^*}>\sqrt{n}/2$ then we are again done. Otherwise we continue
by defining $a_3$ and $c_3$ and so on.  After a finite number of steps, say $m\leq
n+\sqrt{n}/2-N$, we do find $N-k_{j^*}>\sqrt{n}/2$. At each step $l$
we have the inequality \eqref{inductie stap main} with $a$ replaced
by $a_l$ and $a_1$ replaced by $a_{l+1}$ and $j^*$ is with respect
to $a_l$. We can reduce all the inequalities together to the single
inequality
\begin{align*}
\left| \sum_{s=-\sqrt{n}}^{\sqrt{n}} \big({a^{-1}}\big)_{s}
\big({a}\big)_{N-s}-\sum_{s=-\sqrt{n}}^{\sqrt{n}}
\big({a^{-1}_m}\big)_{s} \big({a_m}\big)_{N-s}\right|\leq \frac{D_1
A_1 }{n}.
\end{align*}
Combining this inequality with the fact that the above arguments
show that \eqref{eq: cancellation lemma} holds for $a_m$ leads to
the statement.

Hence it remains to prove (\ref{inductie stap main}). First note
that
\begin{align} \label{decompositie in I123}
\left|  \sum_{s=-\sqrt{n}}^{\sqrt{n}} \big({a^{-1}}\big)_{s}  \big({a}\big)_{N-s}-\sum_{s=-\sqrt{n}}^{\sqrt{n}} \big({a^{-1}_1}\big)_{s}  \big({a_1}\big)_{N-s}\right|\leq I_1 +I_2 +I_3,
\end{align}
where
\begin{align*}
I_1&=\left|  \sum_{s=-\sqrt{n}}^{\sqrt{n}} \big({a^{-1}}\big)_{s}  \big({a}\big)_{N-s}-\sum_{s=-\sqrt{n}}^{\sqrt{n}} \big({a^{-1}_1}\big)_{s}  \big({a}\big)_{N-s}\right| =\left|\sum_{s=-\sqrt{n}}^{\sqrt{n}} \big({a^{-1}_1(c_1^{-1}-1)}\big)_{s}  \big({a}\big)_{N-s}\right|,\\
I_2&=\left|  \sum_{s=-\sqrt{n}}^{\sqrt{n}} \big({a^{-1}_1}\big)_{s} \big({a}\big)_{N-s}-\sum_{s=-\sqrt{n}}^{\sqrt{n}} \big({a^{-1}_1}\big)_{s}  \big({a}_1(1+\log c_1) \big)_{N-s}\right| \\& =\left|\sum_{s=-\sqrt{n}}^{\sqrt{n}} \big({a^{-1}_1}\big)_{s}  \big({a_1 (c_1-1-\log c_1)}\big)_{N-s} \right|,
\end{align*}
and
\begin{align*}
  I_3 =\left|\sum_{s=-\sqrt{n}}^{\sqrt{n}} \big({a^{-1}_1}\big)_{s}  \big({a_1 \log c_1}\big)_{N-s}\right|.
\end{align*}
The terms $I_1$ and $I_2$ can be estimated by the  Cauchy-Schwarz inequality,
\begin{align*}
I_1\leq \|a^{-1}_1( c_1^{-1}-1)\|_{\Ltwee} \left(\sum_{s=-\sqrt{n}}^{\sqrt{n}} \Big|\big({a}\big)_{N-s}\Big|^2\right)^{1/2}
 \leq \frac{\|a^{-1}_1\|_W\| c_1^{-1}-1\|_W  \|a\|_{B^{1/2}_2} }{\sqrt{N-\sqrt{n}} },
\end{align*}
and
\begin{align*}
  I_2&\leq \|a^{-1}_1\|_2 \|a_1(c_1-1-\log c_1)\|_2 \leq \|a^{-1}_1\|_W \|a_1\|_W \|c_1-1-\log c_1\|_W.
\end{align*}
Note that $\|a_1\|_W, \|a_1^{-1}\|_W$ and $\|a\|_{B^{1/2}_2}$ are
all uniformly bounded in $n$ and $N$. Now
\begin{align*}
  \|c_1^{-1}-1\|_W &\leq  \exp\left(\frac{|\alpha_{j^*}|+|\alpha_{-j^*}|}{\sqrt{k_{j^*}}}\right)-1,\\ \|c_1-1-\log c_1\|_W & \leq  \exp\left(\frac{|\alpha_{j^*}|+|\alpha_{-j^*}|}{\sqrt{k_{j^*}}}\right)-1-\frac{|\alpha_{j^*}|+|\alpha_{-j^*}|}{\sqrt{k_{j^*}}}.
\end{align*}
Since $k_{j^*}>n/2$ it follows that
\begin{equation}\label{estimates on I12}
I_{1,2}\leq (|\alpha_{j^*}|+|\alpha_{-j^*}|)D_2
n^{-1},\end{equation}
 for some
constant $D_2$.

This brings us to the most important part of the proof, namely
estimating $I_3$. Note that $\log
c_1=(a_{j^*}z^{k_{j^*}}+\alpha_{-j^*}
z^{-k^{j^*}})/\sqrt{k_{j^{*}}}$. Write
\[  I_3\leq I_{31}+I_{32},\]
where
\begin{align*}
 I_{31}&=
 \left| \frac{\alpha_{k_{j^{*}}}}{\sqrt{k_{j^*}}}   \sum_{s=-\sqrt{n}}^{\sqrt{n}}  \big({a^{-1}_1}\big)_{s} \big({a_1}\big)_{N-s-k_{j^*}}\right|,\qquad I_{32}= \left|\frac{\alpha_{-k_{j^{*}}}}{\sqrt{k_{j^*}}}\sum_{s=-\sqrt{n}}^{\sqrt{n}} \big({a^{-1}_1}\big)_{s}  \big({a_1}\big)_{N-s+k_{j^*}}\right|.
\end{align*}
The term $I_{32}$ can again be estimated  by the Cauchy-Schwarz
inequality. The result is that
\begin{equation}\label{estimate on I32}
  I_{32}\leq  \frac{\sqrt{2} |\alpha_{-k_{j^*}}| \|a_1^{-1}\|_{\Ltwee} \|a_1\|_{B^{1/2}_2}}{\sqrt{k_{j^*} n}} \leq \frac{2 |\alpha_{-k_{j^*}}| \|a_1^{-1}\|_{\Ltwee} \|a_1\|_{B^{1/2}_2}}{n},
\end{equation}  where we used   the fact that
$N-s+k_{j^*}\geq n/2$ if $|s|\leq \sqrt{n}$  and $k_{j^*}\geq n/2$ .

The term $I_{31}$  is more subtle. Since $N>k_{j^*}$ we find
\begin{align*}
\sum_{s=-\sqrt{n}}^{\sqrt{n}}  \big({a^{-1}_1}\big)_{s} \big({a_1}\big)_{N-s-k_{j^*}}
&+\sum_{|s|>\sqrt{n}}  \big({a^{-1}_1}\big)_{s} \big({a_1}\big)_{N-s-k_{j^*}} =\sum_{s}  \big({a^{-1}_1}\big)_{s} \big({a_1}\big)_{N-s-k_{j^*}}\\&=({a^{-1}_1a_1})_{N-k_{j^*}}=0.\end{align*}
Therefore
\begin{align*}
  I_{31}=\frac{|a_{{j^*}}|}{\sqrt{k_{j^*}}} \left|\sum_{|s|>\sqrt{n}}  \big({a^{-1}_1}\big)_{s}
  \big({a_1}\big)_{N-s-k_{j^*}}\right|.
\end{align*}
Now we estimate the latter expression by the Cauchy-Schwarz
inequality again. Note that $|N-s-k_{j^*}|>\sqrt{n}/2$ if
$|s|>\sqrt{n}$. The result is that
\begin{align} \label{estimate on I31}
  I_{31}\leq \frac{2|\alpha_{j^*}| \|a^{-1}_1\|_{B^{1/2}_2}\|a_1\|_{B^{1/2}_2}}{n
  \sqrt{k_{j^*}}}\leq  \frac{2^{3/2}|\alpha_{j^*}|
  \|a^{-1}_1\|_{B^{1/2}_2}\|a_1\|_{B^{1/2}_2}}{n^{3/2}},
\end{align}
where we also used $k_{j^*}\geq n/2$. So from \eqref{estimates on
I12}, \eqref{estimate on I32} and \eqref{estimate on I31} we find
\begin{equation}\label{estimate on I123} I_{1,2,3}\leq
(|a_{k_{j^*}}|+|\alpha_{-k_{j^*}}|) D_3/n\end{equation} for some
constant $D_3$. Now (\ref{inductie stap main}) follows by
(\ref{decompositie in I123}) and \eqref{estimate on I123}. This
proves the statement.
\end{proof}

Now we can prove the following corollary by fairly direct estimates.

\begin{corollary} \label{corr Tr Tn Fn} We have that
\begin{equation}
\Tr T_n(a^{-1})P_n H(a) H(\widetilde{g^{(2)})}) P_n= \OO(n^{-1/4}),
\end{equation}
for $n\to \infty$.
\end{corollary}
\begin{proof}
A straightforward calculation leads to
\begin{align} \label{eq:cor:TrTnFn:expandtrace}
\Tr T_n(a^{-1})P_n H(a) H(\widetilde{g^{(2)})}) P_n=\sum_{k_j > n}
\frac{\alpha_{-j}}{\sqrt{n}}\sum_{s=-n}^n ({a^{-1}})_s ({a})_{k_j-s}
(n-|s|).
\end{align}
We estimate each term in the sum with respect to $k_j$ separately.
So let $k_j>n$. Write
\begin{align}
\sum_{s=-n}^n ({a^{-1}})_s
({a})_{k_j-s}(n-|s|)&=\sum_{s=-\sqrt{n}}^{\sqrt{n}} ({a^{-1}})_s
({a})_{k_j-s} (n-|s|)\nonumber \\
&+\sum_{\sqrt{n}<|s|\leq n}  ({a^{-1}})_s
({a})_{k_j-s} (n-|s|). \label{eq:cor:TrTnFn:splitting}
\end{align}
After some preparation, the rightmost sum of the right-hand side of \eqref{eq:cor:TrTnFn:splitting}
can be estimated by the Cauchy-Schwarz inequality as before
\begin{align}
\Big|\sum_{\sqrt{n}<|s|\leq n} & ({a^{-1}})_s ({a})_{k_j-s}
(n-|s|)\Big| \leq \sum_{\sqrt{n}<|s|\leq n} | ({a^{-1}})_s
({a})_{k_j-s}||n-s|\nonumber \\
& =  \sum_{\sqrt{n}<|s|\leq n}
\sqrt{|s|}|({a^{-1}})_s| \,  \sqrt{k_j-s}\,   |({a})_{k_j-s}|
\,\frac{(n-|s|)}{\sqrt{|s|(k_j-s)}}\nonumber \\
& \leq {\|a^{-1}\|_{B^{1/2}_2} \|a\|_{B^{1/2}_2}}n^{1/4}, \label{eq:cor:TrTnFn:splitting3a}
\end{align}
where we used that
\[\frac{n-|s|}{\sqrt{|s| (k_{j}-s)} }\leq
\sqrt{\frac{n-|s|}{|s|}}\leq n^{1/4},
 \] for all
$\sqrt{n} \leq |s| \leq n$.

Now consider the left sum of the right-hand side of \eqref{eq:cor:TrTnFn:splitting}.
\begin{align}\label{eq:cor:TrTnFn:splitting2}
\Big|\sum_{s=-\sqrt{n}}^{\sqrt{n}} ({a^{-1}})_s ({a})_{k_j-s} (n-|s|)\Big|&=n \Big|\sum_{s=-\sqrt{n}}^{\sqrt{n}} ({a^{-1}})_s ({a})_{k_j-s} \Big|+\Big|\sum_{s=-\sqrt{n}}^{\sqrt{n}} |s|({a^{-1}})_s ({a})_{k_j-s} \Big|.
\end{align}
The second sum of the right-hand side of \eqref{eq:cor:TrTnFn:splitting2}
can again be estimated by a Cauchy-Schwarz argument, from which it follows that it is of order
$n^{-1/2}$. The first sum of the right-hand side of \eqref{eq:cor:TrTnFn:splitting2}  can be dealt with
by  using Lemma \ref{cancellation lemma} and therefore
\begin{align} \label{eq:cor:TrTnFn:splitting3b}
\Big|\sum_{s=-\sqrt{n}}^{\sqrt{n}} ({a^{-1}})_s ({a})_{k_j-s} (n-|s|)\Big|=\OO(n^{1/4}),
\end{align}
for $n\to \infty$.

Inserting \eqref{eq:cor:TrTnFn:splitting3a} and \eqref{eq:cor:TrTnFn:splitting3b} in  \eqref{eq:cor:TrTnFn:splitting}
and using \eqref{eq:cor:TrTnFn:expandtrace} gives
\begin{align*}
\Tr T_n(a^{-1})P_n H(a) H(\widetilde{g^{(2)})}) P_n=\OO(n^{-1/4})\sum_{k_j > n} \alpha_{-j}=\OO(n^{-1/4}),
\end{align*}
for $n\to \infty$. This proves the statement.
\end{proof}
We are almost at the end of our proof. The final thing we need to
show is that the dominant term in $\Tr T_n(a) F_n$ comes from $\Tr
T_n(a^{-1})P_n H(a) H(\widetilde{g^{(2)})}) P_n$, which is small by
the previous corollary.
\begin{corollary} \label{cor:final}
\begin{equation}
\lim_{n \to \infty} \Tr T_n(a^{-1}) F_n=0.
\end{equation}
\end{corollary}
\begin{proof}
Since $W_n^2=P_n$ and by (\ref{Wn rel ctilde}) we find
\begin{align*}
 \Tr T_n(a^{-1}) F_n&= \Tr T_n(a^{-1})P_n H(a) H(\tilde b )P_n + \Tr T_n(a^{-1})W_n H(\tilde a) H(b) W_n\\
&= \Tr T_n(a^{-1})P_n H(a) H(\tilde b )P_n + \Tr W_n  T_n(\widetilde {a^{-1}})P_n H(\tilde a) H(b) W_n\\
&= \Tr T_n(a^{-1})P_n H(a) H(\tilde b )P_n + \Tr  T_n(\widetilde
{a^{-1}})P_n H(\tilde a) H(b) P_n.
\end{align*}
We will only show that $ \Tr T_n(a^{-1})P_n H(a) H(\tilde b )P_n \to
0$. The right term tends to $0$ by  the same arguments. Write
\begin{align}
 T_n(a^{-1})P_n H(a) H(\tilde b )P_n&=\Tr T_n(a^{-1})P_n H(a) H(\tilde b - \widetilde{ g^{(2)}}-1 )P_n\nonumber \\
 &+\Tr T_n(a^{-1})P_n H(a) H(\widetilde{g^{(2)}}
 )P_n. \label{eq:cor:final}
\end{align}
Since
\begin{align*}
\| H(\tilde b - \widetilde{ g^{(2)}}-1 )P_n\|_2\leq
\sqrt{n}\|b-g^{(2)}-1\|_{\Ltwee}\leq \sqrt{n}\left({\rm
e}^{\|g^{(2)}\|_W} -\|g^{(2)}\|_W-1\right),
\end{align*}
and $\|g^{(2)}\|_W\leq A_1/\sqrt{n}$  it follows that
\begin{align*}
|\Tr T_n(a^{-1})P_n & H(a) H(\widetilde b - \widetilde{ g^{(2)}}-1
)P_n|\leq  \| T_n(a^{-1})P_n H(a) H(\tilde b - \widetilde{
g^{(2)}}-1 )P_n\|_1\\& \leq \| T_n(a^{-1})\|_\infty\| P_n H(a)\|_2\|
H(\tilde b - \widetilde{ g^{(2)}}-1 )P_n\|_2=\OO(n^{-\h}),
\end{align*}
for $n\to \infty$. By combining this with \eqref{eq:cor:final} we
see that it only remains to estimate $\Tr T_n(a^{-1})P_n H(a)
H(\widetilde{g^{(2)}} )P_n$, which was done in Corollary \ref{corr
Tr Tn Fn}. This proves the statement.
\end{proof}
Now \eqref{mian result deel2} follows from Corollary \ref{cor:final}
and \eqref{eq:towardsfinal}.

\section{Proof of Theorem \ref{limiet charact functie}} \label{Approximation argument}

We will now show  how the condition $\sum |\alpha_j|<\infty$ can be
made obsolete when we assume that $\alpha_{-j}=\overline{\alpha_j}$.
Let $m\in \N$. We split $X_n$ into two parts
\begin{equation}X_n=X_{n,m}+Y_{n,m}= \sum_{0<|j|\leq m } \frac{\alpha_j}{\sqrt{\min (|k_j|,n)}} \Tr U^{k_j}+
\sum_{|j|> m } \frac{\alpha_j}{\sqrt{\min (|k_j|,n)}} \Tr U^{k_j}.
\end{equation}
Since both $X_{n,m}$ and $Y_{n,m}$ are real we find that
\begin{align}
\left|\E [{\rm e}^{{\rm i} X_n }]-\E[{\rm e}^{ {\rm i} X_{n,m}}]
\right|& = \left|\E[{\rm e}^{ {\rm i} (X_{n,m}+Y_{n,m})}-\E[{\rm
e}^{ {\rm i} X_{n,m}}] \right|\leq   \E[\left|{\rm e}^{ {\rm i}
Y_{n,m}}-1\right|] \nonumber \\&\leq \E[\left| Y_{n,m}\right|] \leq
\E[\left| Y_{n,m}\right|^2] ^{1/2}=  \left(\sum_{|j|>m} |\alpha_j|^2
\right)^{1/2}.
\end{align}
In the last expression we used the fact that the elements
$\frac{1}{\sqrt{\min(|k_j|,n)}} \Tr U^{k_j}$ are orthonormal with
respect to the Haar measure on $U(n)$. It follows that
\begin{equation}
  \limsup \left|\E[{\rm e}^{ {\rm i} (X_{n,m}+Y_{n,m})}-\E[{\rm e}^{ {\rm i} X_{n,m}}]
  \right|\leq \left(\sum_{|j|>m} |\alpha_j|^2 \right)^{1/2}.
\end{equation}
Since $\sum_{|j|\leq m}  | \alpha_j|<\infty$, it follows by Theorem
\ref{main result} and (\ref{connect expect toepl})  that
\begin{equation}
  \lim_{n\to \infty} \E[{\rm e}^{{\rm i} X_{n,m}}]={\rm e}^{-\sum_{j=1}^{m} |\alpha_j|^2 }.
\end{equation}
Hence
\begin{align}
  \limsup_{n\to \infty} &\left|\E[{\rm e}^{ {\rm i} (X_{n,m}+Y_{n,m})}- {\rm e}^{-\sum_{j=1}^{\infty} |\alpha_j|^2}\right|
  \leq  \limsup_{n\to \infty} \left|\E[{\rm e}^{ {\rm i} (X_{n,m}+Y_{n,m})}- \E[{\rm e}^{{\rm i} X_{n,m}} ]\right|
  \\
&+\limsup_{n\to \infty} \left|\E[{\rm e}^{{\rm i} X_{n,m}}]-{\rm e}^{-\sum_{j=1}^{ m} |\alpha_j|^2 }\right| +
\left|{\rm e}^{-\sum_{j=1}^{m} |\alpha_j|^2}-{\rm e}^{-\sum_{j=1}^{\infty} |\alpha_j|^2}\right|\\
&\leq \left(\sum_{|j| > m } |\alpha_j|^2 \right)^{1/2}+ \left|{\rm
e}^{-\sum_{j=1}^{m} |\alpha_j|^2}-{\rm
e}^{-\sum_{j=1}^\infty |\alpha_j|^2}\right|.
\end{align}
If we let $m\to \infty$ the right-hand side tends to zero.

\section{Some comments on more general $n$-dependence}
\label{Generalisatie and Sosh}

The $n$-dependence in the symbols we consider is of a special type.
Let $U$ be a $n\times n$ unitary matrix randomly chosen with respect to the Haar measure. Consider
the random variable $X_n$ by
\begin{equation}
  X_n(U) =\sum_{|j|> 0} \frac{\alpha_{j}(n)}{\sqrt{\min(|k_j(n)|,n)}} \Tr U^{k_j(n)},
\end{equation}
where $\alpha_j(n)$ now also depends on $n$. Again we assume that for each $n$ we have that $\alpha_j(n)=\overline{\alpha_{-j}(n)}$,
 $\{k_j(n)\}_{j\in \N}$ is a sequence of mutually distinct positive integers
 and $k_{-j}(n)=-k_j(n)$. Define
\begin{equation}
  \sigma_n^2=2 \sum_{j=1}^\infty |\alpha_j(n)|^2,
\end{equation}
and assume that $\sigma_n \to \sigma$ as $n\to \infty$ for some
$\sigma$. A natural question is now under what conditions it is still true that
\begin{equation} \label{lim general}
  \lim_{n\to \infty} \mathbb{E}[{\rm e}^{{\rm i} t X_n}]={\rm e}^{- t \sigma^2/2}.
\end{equation}
Since then $X_n$ converges to a complex normal with mean zero and variance $\sigma^2$.
Although, it is known in some cases that it is true,  it will not hold in general.

We will illustrate the subtleties that are involved by an explicit
example inspired on \cite{Sosh}. Let $f$ be a $C^\infty$ function
with support within $[-\pi,\pi]$ and let $0<\gamma\leq 1$. Define
$k_{j}(n)=j$ and
\begin{equation}
  \alpha_{j}(n)=\frac{\sqrt{\min(|j|,n)}}{2\pi n^\gamma} \hat f (j/n^\gamma),
\end{equation}
for all $j$ and $n$. Here $\hat f$ stands for the Fourier transform
of $f$. We assume that \begin{equation}   \hat f(0)=\int_\R f(x)\
{\rm d} x=0.
\end{equation}The random variable $X_n$ can
now be rewritten as
\begin{equation}
  X_n(U) =\sum_{\mu=1}^n f(n^\gamma \theta_\mu).
\end{equation}
Since $f$ has compact support $X_n$ only depends on a few
eigenvalues, for which $\theta_\mu$ is close to zero. If
$0<\gamma<1$, then it is true that $X_n \to N(0,\sigma^2)$, where
\begin{equation}
  \sigma^2=\frac{1}{4\pi^2}\int |y| |\hat f(y)|^2\ {\rm d} y,
\end{equation}
assuming that the latter is finite. This is proved by Soshnikov
\cite{Sosh}.

However, the result does not longer hold for $\gamma=1$. This case
is considered by Hughes and Rudnick in \cite{HughRud1} and for the
classical compact groups other then $U(n)$ in \cite{HughRud2}. In
these works the authors analyzed the limiting behavior of the
moments $\E(X_n^m)$ for $m\in \mathbb{N}$ and proved that  in
general the limiting value of the moments depend on $f$ and are
certainly not Gaussian moments. Hence a result like (\ref{lim
general}) can not hold. However, if $\supp \hat f \subset
[-2/m,2/m]$ then the $m$-th moment does converge to the $m$-th
moment of the normal distribution with mean zero and variance
\begin{equation}
\sigma^2= \frac{1}{4 \pi^2} \int_{-\infty}^\infty \min(|y|,1) |\hat
f (y)|^2 \ {\rm d } y.
\end{equation}
This phenomenon is called mock-Gaussian behavior in \cite{HughRud1}.

\appendix

\section{Acknowledgements}

The presented work was developed whilst the first author was staying
at the Royal Institute of Technology in Stockholm during the spring
term of 2006. The authors wish to thank Jens Hoppe for inviting the
first author and for his generous hospitality during this period.

The authors also wish to thank Zeev Rudnick for drawing  attention
to the papers \cite{HughRud1} and \cite{HughRud2}.

\bibliographystyle{amsplain}

M. Duits: Department of Mathematics, Katholieke Universiteit Leuven,
Celestijnenlaan 200 B, 3001 Leuven, Belgium

e-mail: maurice.duits@wis.kuleuven.be\\

K. Johansson:  Department of Mathematics,  Royal Institute of
Technology,  SE-100 44 Stockholm, Sweden.

e-mail: kurtj@kth.se
\end{document}